\documentclass[prb,superscriptaddress,twocolumn]{revtex4}

\usepackage{amsmath}
\usepackage{amsfonts}
\usepackage{amssymb}
\usepackage{txfonts}
\usepackage{bm}
\usepackage{tabularx}
\usepackage{graphicx,color}
\usepackage{ulem}
\usepackage{mathrsfs}

\def\Vec#1{\bm{#1}}
\def\Hc2{H_\mathrm{c2}}

\def\piz{{\rm PrIr_2Zn_{20}}}

\begin{document}

\title{
Anisotropy of the magnetic-field-induced phase pocket in the non-Kramers doublet system PrIr$_2$Zn$_{20}$}

\author{Shunichiro Kittaka}
\affiliation{Department of Physics, Faculty of Science and Engineering, Chuo University, Bunkyo, Tokyo 112-8551, Japan}
\affiliation{Institute for Solid State Physics, University of Tokyo, Kashiwa, Chiba 277-8581, Japan}
\author{Takahiro Onimaru}
\affiliation{Department of Quantum Matter, Graduate School of Advanced Science and Engineering, Hiroshima University, Higashi-Hiroshima, Hiroshima 739-8530, Japan}
\author{Keisuke T. Matsumoto}
\affiliation{Graduate School of Science and Engineering, Ehime University, Matsuyama, Ehime 790-8577, Japan}
\author{Toshiro Sakakibara}
\affiliation{Institute for Solid State Physics, University of Tokyo, Kashiwa, Chiba 277-8581, Japan}

\date{\today}

\begin{abstract}

We provide thermodynamic evidence for the presence of a magnetic-field-induced small phase pocket near the antiferro-quadrupole (AFQ) phase boundary in the non-Kramers $\Gamma_3$ doublet system $\piz$.
In particular, we measured the specific heat as functions of temperature $T$, magnetic field $B$, and field angle $\phi_B$, and 
found a second specific-heat anomaly in a relatively wide field-angle range near $B \parallel [001]$, although fine tuning of the field strength is required.
We also investigated the rotational magnetocaloric effect and evaluated an entropy change in this phase pocket. 
The present findings demonstrate that multipole degrees of freedom give rise to a magnetic-field-induced exotic order in $\piz$, suggesting the possibility of switching between the order parameters or emergence of a multiple-$\Vec{q}$ order of quadrupoles.
\end{abstract}

\maketitle

Unveiling novel quantum phases driven by higher-order multipole moments is one of the central issues in condensed matter physics.\cite{Onimaru2016JPSJ} 
Among various systems, Pr-based compounds having the non-Kramers doublet $\Gamma_3$ ground state in the cubic crystalline electric field (CEF) have attracted much interest 
because magnetic dipole moments are absent; only electric quadrupole $O_{20}$, $O_{22}$, and octupole $T_{xyz}$ moments are active.
For instance, the $\Gamma_3$ system PrPb$_3$ exhibits an antiferro-quadrupole (AFQ) order at $T_{\rm Q}=0.4$~K,\cite{Onimaru2005PRL} 
and shows a magnetic-field-induced phase transition, likely from the $O_{20}$ to $O_{22}$ phase, under a magnetic field $B$ along the $[110]$ axis.\cite{Onimaru2007JPSJ,Sato2010JPSJ,Onimaru2007JPCS} 
Thus, the non-Kramers $\Gamma_3$ systems provide great opportunities to study multipole physics.

Recently, numerous experimental efforts have been made on the Pr$T_2X_{20}$ system ($T$: transition metals, $X$: Al, Zn and Cd) which has a cubic CeCr$_2$Al$_{20}$-type structure.
So far, it has been well established that the CEF ground state of some Pr$T_2X_{20}$ compounds is a non-magnetic $\Gamma_3$ doublet, and
a wide variety of multipole phenomena have been found in this system.
For instance, PrTi$_2$Al$_{20}$ exhibits a ferro-quadrupole (FQ) order~\cite{Sakai2011JPSJ} at $T_{\rm FQ}=2$~K and
shows a magnetic-field-induced first-order phase transition driven by change in the FQ order parameter due to the competition between anisotropic quadrupole interaction and Zeeman effect.\cite{Taniguchi2019JPSJ, Kittaka2020JPSJ}
In addition, PrV$_2$Al$_{20}$ exhibits an AFQ order below 0.6~K,\cite{Sakai2011JPSJ} 
possibly accompanied by a double transition related to quadrupole and octupole orderings;\cite{Tsujimoto2014PRL} 
it has a high-field phase associated with a rearrangement of quadrupole moments under $B \parallel [100]$.\cite{Shimura2013JPSJ,Shimura2019PRL} 
Moreover, $\piz$~\cite{Onimaru2011PRL} and PrRh$_2$Zn$_{20}$~\cite{Onimaru2012PRB} show an AFQ order at $T_{\rm Q}=0.11$ and 0.06~K, respectively.
Since these Pr-based materials exhibit superconductivity within the ordered phase,\cite{Sakai2012JPSJ,Matsubayashi2012PRL,Tsujimoto2014PRL,Onimaru2010JPSJ,Onimaru2012PRB} 
multipole fluctuations are expected to play a key role in mediating Cooper pairs.

In this paper, we focus on the non-Kramers doublet system $\piz$.
Low-temperature properties of this material are mainly dominated by the non-magnetic $\Gamma_3$ doublet
because the entropy release estimated from specific-heat measurements reaches $R\ln 2$ [$\sim 5.76$~J/(mol K)] at 2~K in zero field.
Indeed, the energy gap between the ground state of a $\Gamma_3$ doublet and the first-excited state of a $\Gamma_4$ triplet is estimated to be 30~K.\cite{Onimaru2011PRL} 
However, at $T_{\rm Q}=0.11$~K, the entropy release is only 20\% of $R\ln 2$.
Since the non-Fermi-liquid (NFL) nature has been observed above $T_{\rm Q}$, 
the rest of the entropy is likely released by the formation of a quadrupole Kondo lattice due to the hybridization between quadrupoles and conduction electrons.\cite{Onimaru2016PRB} 
The AFQ phase is destroyed by a magnetic field of 5~T along the $[001]$ axis, 
whereas it survives at higher magnetic fields along the $[111]$ and $[110]$ axes.\cite{Ishii2011JPSJ} 
The propagation vector was revealed to be $\Vec{k}=(1/2, 1/2, 1/2)$ from neutron diffraction measurements in magnetic fields of $B \parallel [110]$,\cite{Iwasa2017PRB} 
indicating that the $O_{22}$-type order parameter is dominant.
Although a possible second transition was observed in the field range $1.5\ {\rm T} \le B \le 3$~T for $B \parallel [001]$ in the earlier specific-heat study,\cite{Onimaru2011PRL} 
it was not detected in the recent one.\cite{Onimaru2016PRB} 
Therefore, it was interpreted to be due to an extrinsic effect.\cite{Onimaru2016PRB} 
However, in the present work, we have performed field-angle-resolved measurements of the specific heat and revealed that the double transition is intrinsic and strongly depends on the field strength and its orientation. 

High-quality single crystals of $\piz$ were grown by the melt-growth method.\cite{Onimaru2010JPSJ} 
A single-crystalline sample with its  mass of 1.24~mg was used in this study.
The specific heat $C$ was measured by the standard quasi-adiabatic heat-pulse method using a homemade calorimeter in a dilution refrigerator (Oxford, Kelvinox AST Minisorb).
The sample was attached on the stage of our calorimeter so that the $[1\bar{1}0]$ axis is roughly along the vertical direction.
The magnetic field $B$ was applied by using a vector magnet which generates up to 7~T (3~T) along a horizontal (vertical $z$) direction.
The orientation of the magnetic field can be controlled in three dimensions by rotating the refrigerator around the $z$ axis using a stepping motor.
In this study, we precisely applied the magnetic field along the $(1\bar{1}0)$ plane, 
including the $[001]$, $[112]$, $[111]$, and $[110]$ axes.
From the field-angle dependence of its specific heat, it was found that the $[1\bar{1}0]$ axis of the present sample is tilted away from the $z$ direction by roughly $10^\circ$ due to the non-rectangular sample shape.
Therefore, we tuned the horizontal and vertical magnetic fields at each field angle so that the magnetic field orients to the target direction with high accuracy of better than $0.1^\circ$.
The relative change in the entropy with rotating the magnetic field was investigated using the rotational magnetocaloric technique.\cite{Kittaka2018JPSJ} 

\begin{figure}
\includegraphics[width=3.4in]{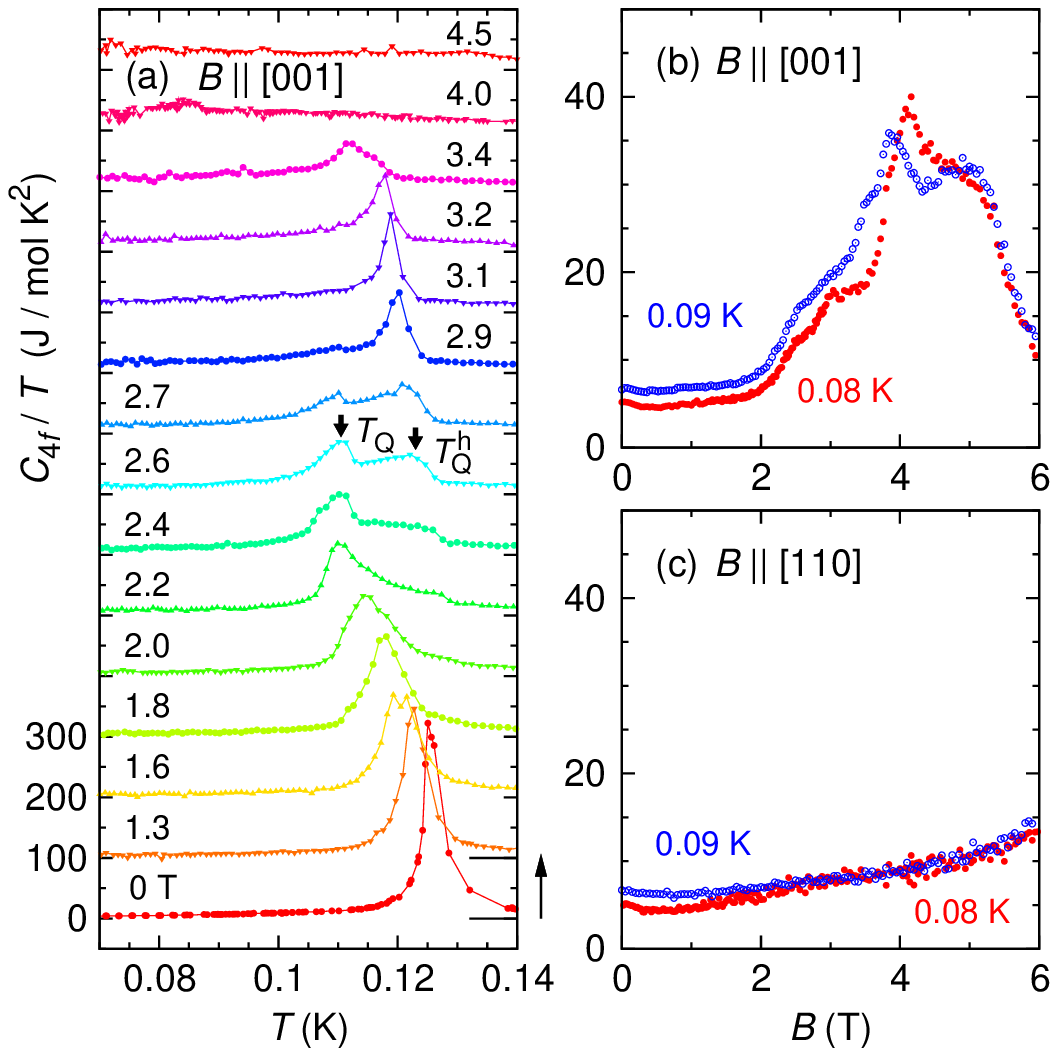}
\caption{
(a) Temperature dependence of the $4f$ contribution to the specific heat of $\piz$, $C_{4f}/T$, in various magnetic fields applied along the $[001]$ direction. 
Each set of the data is shifted vertically by 100 J/(mol K$^2$) for clarity.
(b), (c) Field dependence of $C_{4f}/T$ at 0.08 and 0.09~K along the (b) $[001]$ and (c) $[110]$ axes. 
}
\label{CT}
\end{figure}

Figure \ref{CT}(a) shows the temperature dependence of the $4f$ contribution to the specific-heat data, $(C-C_{\rm N})/T$, hereafter referred to as $C_{4f}/T$, 
measured under a magnetic field applied along the cubic $[001]$ axis.
Here, the nuclear contribution to the specific heat, $C_{\rm N}$, is calculated by using the Hamiltonian of a nuclear spin of a Pr nucleus ($I=5/2$) [see details in the Supplemental Material (SM) of Ref.~\onlinecite{Kittaka2020JPSJ}].
For simplicity, in this calculation, the site-averaged magnitude of the Pr magnetic moment is assumed to be $m_{\rm Pr}(B)= 0.202B$ $\mu_{\rm B}/$Pr below 6~T;\cite{Onimaru2016PRB} 
the field-angle dependence of $m_{\rm Pr}(B)$ is neglected although the actual magnetization is weakly anisotropic.\cite{Onimaru2011PRL} 
In zero field, the present sample exhibits a sharp specific-heat peak at $T_{\rm Q}=0.125$~K.
This peak is much sharper and $T_{\rm Q}$ is slightly higher compared with the previous reports~\cite{Onimaru2011PRL, Onimaru2016PRB} [see SM (I)~\cite{SM}]. 

With increasing the magnetic field along the $[001]$ axis, $T_{\rm Q}$ decreases monotonically up to 2~T, consistent with the previous report.\cite{Onimaru2016PRB} 
In the magnetic-field range between 2 and 3~T, $T_{\rm Q}$ remains approximately 0.11~K, independent of the magnetic-field strength.
However, the specific-heat peak at $T_{\rm Q}$ is suppressed with increasing $B$, and it is smeared out around 3~T.
Instead, a second specific-heat anomaly develops at a slightly higher temperature of $T_{\rm Q}^{\rm h}\sim0.12$ K above 2 T.
These two specific-heat anomalies clearly coexist in the field range 2.2~T $\lesssim B \lesssim$ 2.7~T, as shown in Fig.~\ref{CT}(a).
This second peak at $T_{\rm Q}^{\rm h}$ becomes most remarkable at 3.1~T and becomes broader at higher $B$.

In the earlier specific-heat study,\cite{Onimaru2011PRL} a similar double-peak structure was observed in the field range $1\lesssim B \lesssim 3$~T using a sample whose specific heat shows a relatively broad transition at $T_{\rm Q}$ of 0.11~K in zero field.
However, it was not detected in a more recent study,\cite{Onimaru2016PRB} which reports the temperature dependence of the specific heat at 1~T intervals, using a different sample with a slightly higher $T_{\rm Q}$ [see SM (I)~\cite{SM}].
These facts imply that the double-peak structure depends on the sample quality and that fine tuning of the magnetic field is essential to observe it.

Figure~\ref{CT}(b) shows the field dependence of $C_{4f}/T$ for $B \parallel [001]$ at 0.08 and 0.09~K.
Above 2~T, the low-temperature $C_{4f}(B)$ increases remarkably with $B$ up to 4~T.
A clear peak at $B \sim 4$~T indicates the development of an exotic Fermi-liquid (FL) state, as reported previously.\cite{Onimaru2016PRB} 
A kink anomaly at 3~T might be related to the disappearance of the specific-heat peak at $T_{\rm Q}$; 
a corresponding anomaly was likely to have been observed in the elastic moduli $C_{11}$,\cite{Ishii2011JPSJ} the resistivity~\cite{Ikeura2017JPSCP} and its coefficient $A$.\cite{Onimaru2016PRB} 
No prominent anomaly was found in $C_{4f}(B)$ for $B \parallel [110]$, as shown in Fig.~\ref{CT}(c). 

\begin{figure}
\includegraphics[width=3.4in]{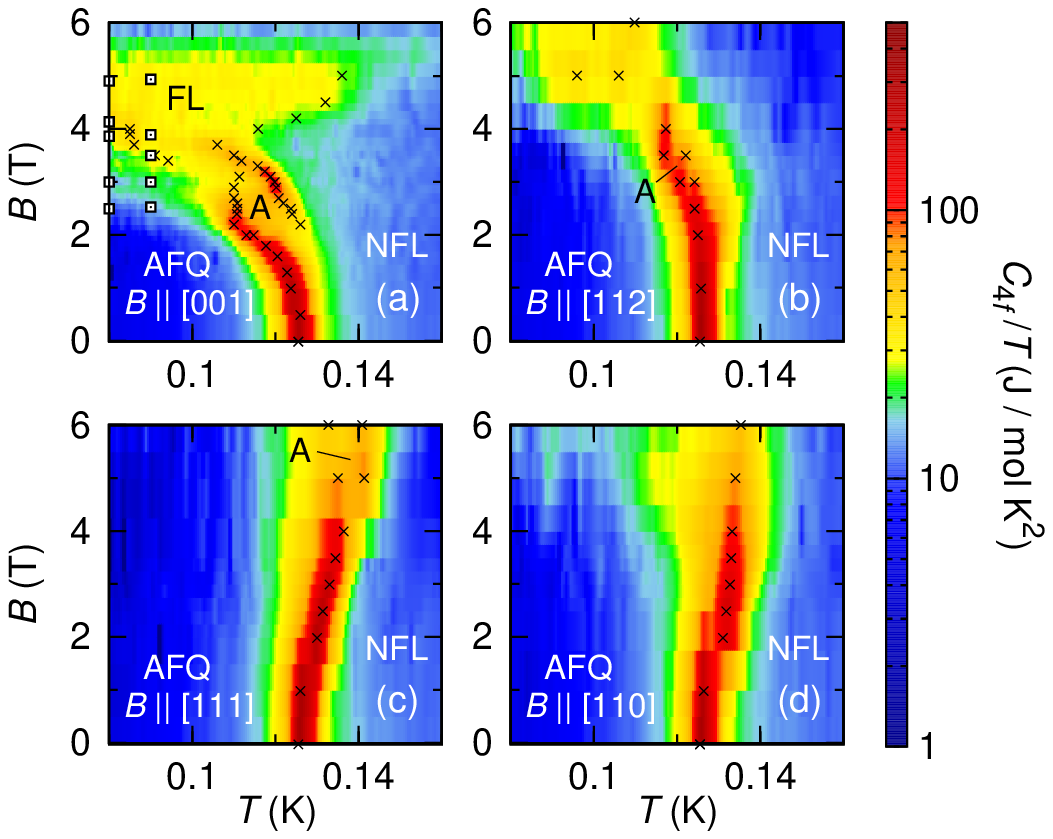}
\caption{
Contour plots of $C_{4f}/T$ in the $B$--$T$ plane for magnetic-field orientations along the (a) $[001]$, (b) $[112]$, (c) $[111]$, and (d) $[110]$ axes.
Crosses (squares) represent the peak or shoulder anomalies observed in $C_{4f}(T)/T$ [$C_{4f}(B)$].
}
\label{BT}
\end{figure}

The $B$--$T$ phase diagram for $B \parallel [001]$, along with a contour plot of $C_{4f}(B,T)/T$, is shown in Fig.~\ref{BT}(a).
Here, the specific-heat anomalies found in the temperature scan are represented by crosses.
The presence of a magnetic-field-induced phase pocket, hereafter referred to as ``the A phase'', is suggested by the double-peak structure in the $C(T)$ data.
This A phase occurs at higher temperatures than $T_{\rm Q}(B)$.
It is unclear whether or not $T_{\rm Q}(B)$ and $T_{\rm Q}^{\rm h}(B)$ lines merge because either specific-heat anomaly is smeared out near possible merging points.

Temperature dependences of $C_{4f}/T$ in different field orientations for $B \parallel [112]$, $[111]$, and $[110]$ have also been investigated [see SM (II)~\cite{SM}].
Figures \ref{BT}(b)-(d) display contour plots of $C_{4f}/T$ in the $B$--$T$ plane for $B \parallel [112]$, $[111]$, and $[110]$, respectively.
Whereas the magnetic-field-induced double transition can be seen for $B \parallel [112]$ and $[111]$, it is absent in $B \parallel [110]$ at least below 6~T.

\begin{figure}
\includegraphics[width=3.4in]{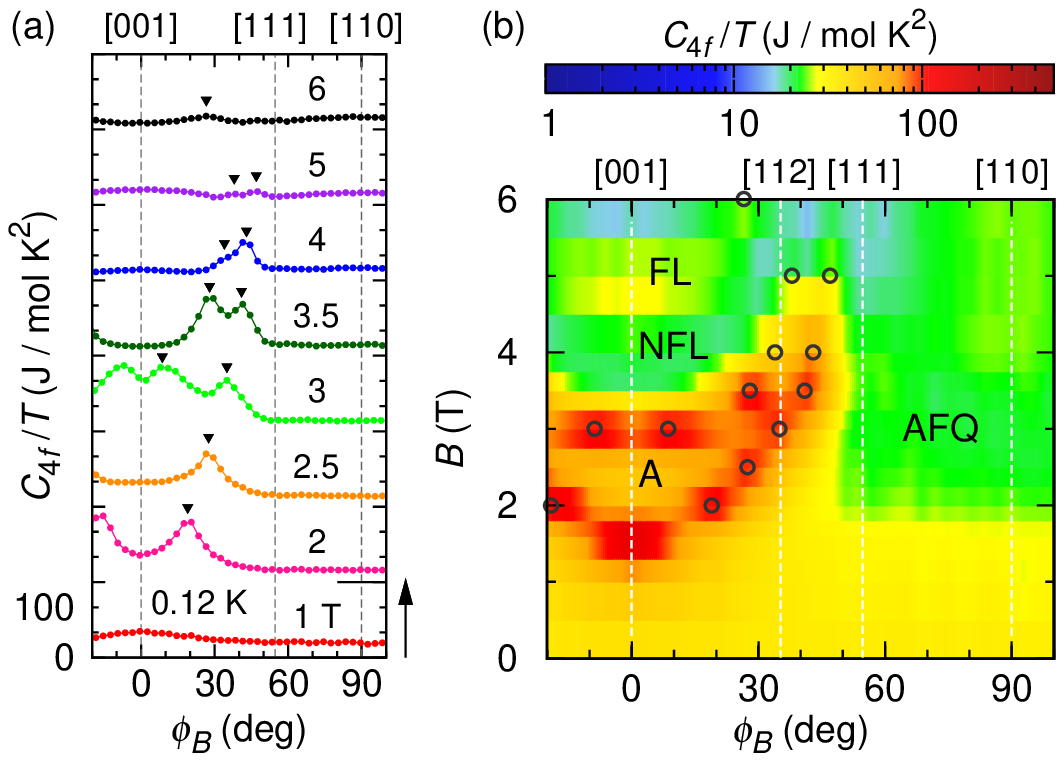}
\caption{
(a) Field-angle dependence of $C_{4f}/T$ at 0.12~K under a magnetic field rotated within the $(1\bar{1}0)$ plane.
Numbers labeling the curves represent the rotating magnetic field in Tesla.
Each set of the data is shifted vertically by 150 J/(mol K$^2$) for clarity.
(b) Contour plot of $C_{4f}/T$ at 0.12~K in the $B$--$\phi_B$ plane.
Circles in (b) show the location of the peak anomalies represented by triangles in (a).
}
\label{Bphi}
\end{figure}

To clarify the field-orientation dependence of the A phase, 
we measured $C_{4f}$ at several magnetic fields rotated within the $(1\bar{1}0)$ plane.
The results of $C_{4f}(\phi_B)$ at 0.12~K are shown in Fig.~\ref{Bphi}(a),
where $\phi_B$ is the field angle measured from the $[001]$ axis within the $(1\bar{1}0)$ plane. 
At 3 and 3.5~T, two peaks can be clearly seen in $C_{4f}(\phi_B)$. 
Although the peak intensity is significantly suppressed above 4~T, the double-peak feature can be confirmed even at 5~T. 
These peaks are represented by circles in a contour plot of $C_{4f}(B,\phi_B)/T$ at 0.12~K [Fig.~\ref{Bphi}(b)]. 
The A phase exists in a relatively wide $\phi_B$ range and shifts toward higher magnetic fields with increasing $\phi_B$ from $0^\circ$.
This symmetric field-angle dependence eliminates the possibility that 
the second transition is caused by a domain with a tilted crystalline axis in the sample.

\begin{figure}
\includegraphics[width=3.4in]{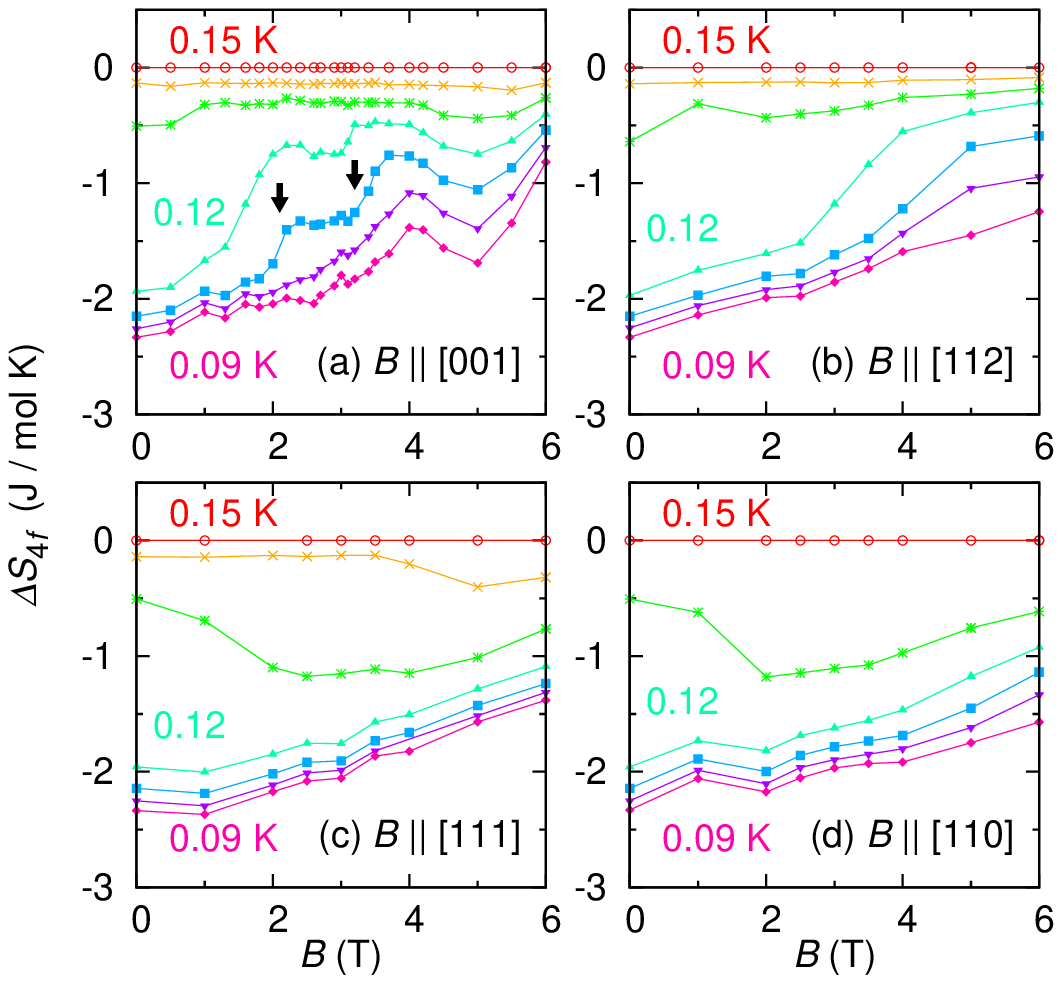}
\caption{
The relative change in the entropy measured from the value at 0.15~K, $\Delta S_{4f}(T)$, at 0.15, 0.14, 0.13, 0.12, 0.11, 0.10, and 0.09 K (from top to bottom),
plotted as a function of $B$ along the (a) $[001]$, (b) $[112]$, (c) $[111]$, and (d) $[110]$ axes.
}
\label{SB}
\end{figure}

In order to obtain the entropy information, 
the temperature dependence of the entropy relative to the value at 0.15~K was estimated as 
\begin{equation}
\Delta S_{4f}(T)=S_{4f}(T)-S_{4f}(0.15\ {\rm K})=-\int_T^{0.15\ {\rm K}} \frac{C_{4f}(T)}{T}dT
\end{equation}
for each magnetic field [see SM (II)~\cite{SM}].
Then, the data points of $\Delta S_{4f}$ at several selected temperatures were extracted and plotted as a function of $B$, $\Delta S_{4f}(B)$, in Figs.~\ref{SB}(a)-\ref{SB}(d)
for each field orientation.
From $\Delta S_{4f}(B)=S_{4f}(T^\ast,B)-S_{4f}(0.15\ {\rm K},B)$ at a selected temperature $T^\ast$,
we can detect the field variation of $S_{4f}(T^\ast,B)$ itself when $S_{4f}(0.15\ {\rm K},B)$ is independent of $B$. 
In the previous study,\cite{Onimaru2016PRB} it was revealed that, for $B \parallel [001]$, 
$S_{4f}(0.15\ {\rm K},B)$ is nearly unchanged up to 3~T, but it is suppressed above 5~T due to the increase in the characteristic temperature $T_0$, which is related to the formation of the quadrupole Kondo lattice.
Because $T_0$ estimated from resistivity measurements is nearly constant up to 4~T for $B\parallel [001]$,\cite{Onimaru2016PRB} 
one might assume that the present base entropy $S_{4f}(0.15\ {\rm K},B)$ is independent of $B$ up to 4~T for $B \parallel [001]$.

At 0.11-0.12~K, an apparent $\Delta S_{4f}(B)$ plateau was observed in the field range $2\ {\rm T} \lesssim B \lesssim 3$~T, bounded by the two arrows in Fig.~\ref{SB}(a);
$\Delta S_{4f}(B)$ for $B \parallel [001]$ at 0.11~K exhibits rapid enhancements at the boundaries. 
In addition, the magnetic-field-induced transition becomes sharpest at 3.1~T for $B \parallel [001]$ [Fig.~\ref{CT}(a)],
and the specific heat anomalies disappear where the phase transition terminates. 
Further investigations, such as high-resolution magnetostriction \cite{Worl2019PRB,Kittaka2023PRB} and magnetization measurements with fine tuning of temperature, are required to clarify whether the phase transition to the A phase is of first order.

In the previous reports,\cite{Onimaru2011PRL,Onimaru2016PRB} the entropy release below $T_{\rm Q}$ was estimated to be 2 J/(mol K) in zero field.
As presented in Figs.~\ref{SB}(a)-\ref{SB}(d), 
$|\Delta S_{4f}(B)|$ at 0.09~K (well below $T_{\rm Q}$) remains roughly 2~J/(mol K) up to 3~T in any field orientation;
no prominent entropy change was observed around 2~T for $B \parallel [001]$, 
at which the elastic modulus $C_{11}$ (Ref.~\onlinecite{Ishii2011JPSJ}) and $C_{4f}(B)$ [Fig.~\ref{CT}(b)] show an anomaly.
This fact supports the absence of the phase boundary at 2~T or tiny entropy change at this possible phase boundary.

\begin{figure}
\includegraphics[width=3.4in]{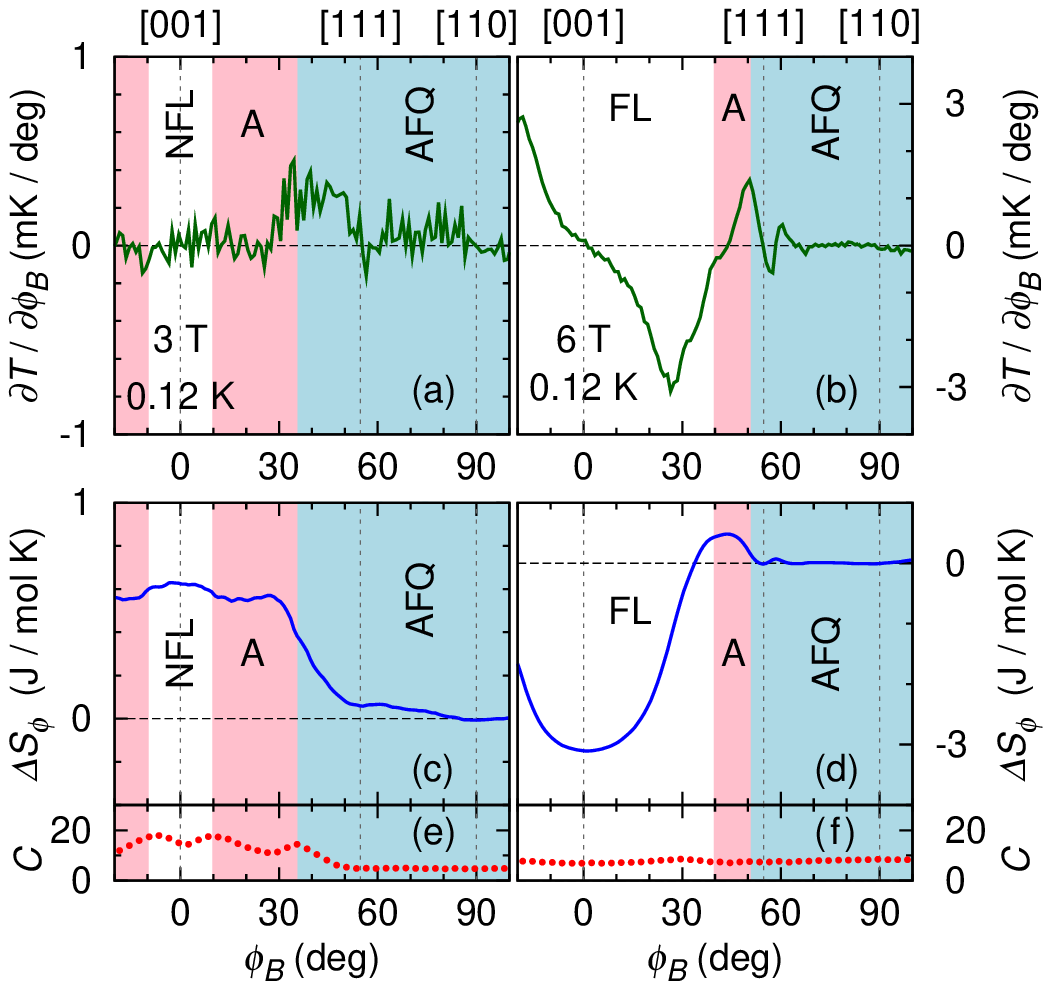}
\caption{
Field-angle dependences of (a), (b) the rotational magnetocaloric effect $(\partial T/\partial \phi_B)_S$, (c), (d) the relative change in the entropy $\Delta S_\phi$, and (e), (f) the specific heat $C$ 
under a magnetic field of 3 and 6 T, respectively, rotated within the $(1\bar{1}0)$ plane at 0.12~K.
Here, $C$ and $\Delta S_\phi$ contain nuclear contribution.
}
\label{Sp}
\end{figure}

To clarify the field-angle dependence of the entropy, the rotational magnetocaloric effect, $(\partial T/\partial \phi_B)_S$, was investigated at 0.12~K
under a magnetic field of 3 and 6~T rotated within the $(1\bar{1}0)$ plane; 
the results are shown in Figs.~\ref{Sp}(a) and \ref{Sp}(b), respectively.  
Combined with the $C(\phi_B)$ data at 0.12~K in Figs. \ref{Sp}(e) and \ref{Sp}(f), 
the $\phi_B$ dependence of the relative change in the entropy has been investigated as 
\begin{equation}
\Delta S_\phi=S(\phi_B)-S(90^\circ)=-\int_{90^\circ}^{\phi_B}\frac{C}{T}\biggl(\frac{\partial T}{\partial\phi_B}\biggl)_S d\phi_B.
\end{equation}
The results at 3 and 6~T are shown in Figs. \ref{Sp}(c) and \ref{Sp}(d), respectively.
The $\Delta S_{4f}(B)$ plateau in the A phase can be again confirmed from $\Delta S_\phi$ at 0.12~K and 3~T in the angle range $10^\circ\lesssim \phi_B \lesssim 30^\circ$ [Fig~\ref{Sp}(c)].
The phase transition from the NFL state to the A phase at 0.12 K reduces the entropy by only 0.07 J/(mol K), much less than the entropy decrease of roughly 0.5 J/(mol K) in the phase transition from the A phase to the AFQ phase.
These results along with the $\Delta S_\phi$ plateau suggest that the ground state of the A phase is well separated from the first-excited state
and still has a substantial entropy.

At 0.12~K and 6~T, $\Delta S_\phi$ reaches its largest value of 0.48~J/(mol K) at $\phi_B \sim 40^\circ$, 
which is comparable to the $\Delta S_\phi$ value in the A phase at 3~T.
Therefore, this peak may also be related to the occurrence of the A phase.
However, at high magnetic fields, it is difficult to identify the boundary of the A phase from $C(\phi_B)$ and $S(\phi_B)$,
because $T_0$ drastically depends on the field angle.

Finally, let us discuss possible origins for the A phase in $\piz$.
A first possible scenario is switching of the AFQ order parameters from $O_{22}$ to $O_{20}$.
Indeed, the competition between quadrupole interaction and the Zeeman effect can lead to switching of order parameters, as predicted from theoretical studies.\cite{Hattori2014JPSJ,Hattori2016JPSJ} 
However, it is still uncertain whether this switching can occur only in the high-temperature region;
usual phase transitions understood in terms of a change in the order parameter, as in the case of PrPb$_3$, are accompanied by a change of the ground state at 0~K.
Such a high-temperature phase pocket without changing the ground state at 0 K is unusual for the $\Gamma_3$ doublet system.
A second possible scenario is the occurrence of an exotic multipole ordering such as multiple-$\Vec{q}$ orders of quadrupoles.
The magnetic-field-induced phase pocket in $\piz$ is reminiscent of the skyrmion-lattice phase (so-called A phase) in MnSi, 
which is induced under a magnetic field in a narrow temperature and magnetic-field range just below the helimagnetic transition temperature.
In the A phase of MnSi, a superposition of three helical states, the so-called triple-$\Vec{q}$ state, was observed via small-angle neutron scattering experiments.\cite{Muhlbauer2009Science} 
Similar two specific-heat anomalies were reported in MnSi as well.\cite{Bauer2013PRL} 
Not only in the non-centrosymmetric systems, skyrmion-lattice phases have also been found in the centrosymmetric lattice systems.\cite{Spachmann2021PRB,Hirschberger2019NC,Yasui2020NC, Ishiwata2020PRB} 
On theoretical grounds, rich phase diagrams in the $\Gamma_3$ doublet system, such as triple-$\Vec{q}$ orders of multipoles~\cite{Tsunetsugu2021JPSJ,Ishitobi2021PRB} and the composite state,\cite{Inui2020PRB} have been proposed recently.
Further investigations, such as neutron-scattering experiments and nuclear-magnetic-resonance measurements, are essential to elucidate the nature of the A phase in $\piz$. 

In summary, we have provided thermodynamic evidence for the presence of a magnetic-field-induced multipole ordered phase (so-called the A phase) in $\piz$.
The A phase occurs only in the high-temperature region and depends strongly on the magnetic-field orientation. 
Such a high-temperature phase pocket is novel in the $\Gamma_3$ doublet system and
paves the path to further understanding of the wide variety of multipole orders.

\begin{acknowledgments}
We thank K. Hattori for useful discussion. This work was supported by KAKENHI (23H04868, 23H04870, 23H01128, 18H04306) from JSPS.
\end{acknowledgments}

\clearpage
\onecolumngrid
\appendix

\begin{center}
{\large Supplemental Material for \\
Anisotropy of the magnetic-field-induced phase pocket in the non-Kramers doublet system $\piz$}\\
\vspace{0.1in}
Shunichiro Kittaka,$^{1,2}$ Takahiro Onimaru$^3$, Keisuke T. Matsumoto$^{4}$, and Toshiro Sakakibara$^{2}$\\
{\small 
\textit{$^1$Department of Physics, Faculty of Science and Engineering, Chuo University, Bunkyo, Tokyo 112-8551, Japan}\\
\textit{$^2$Institute for Solid State Physics, University of Tokyo, Kashiwa, Chiba 277-8581 \\}
\textit{$^3$Department of Quantum Matter, Graduate School of Advanced Science and Engineering,\\ Hiroshima University, Higashi-Hiroshima, Hiroshima 739-8530, Japan}\\
\textit{$^4$Graduate School of Science and Engineering, Ehime University, Matsuyama, Ehime 790-8577, Japan\\}
}
(Dated: \today)
\end{center}

\renewcommand{\thefigure}{S\arabic{figure}}
\renewcommand{\thetable}{S\arabic{table}}
\setcounter{figure}{0}

\section*{I. Sample dependence of the specific heat}

\begin{figure}
\begin{center}
\includegraphics[width=6.in]{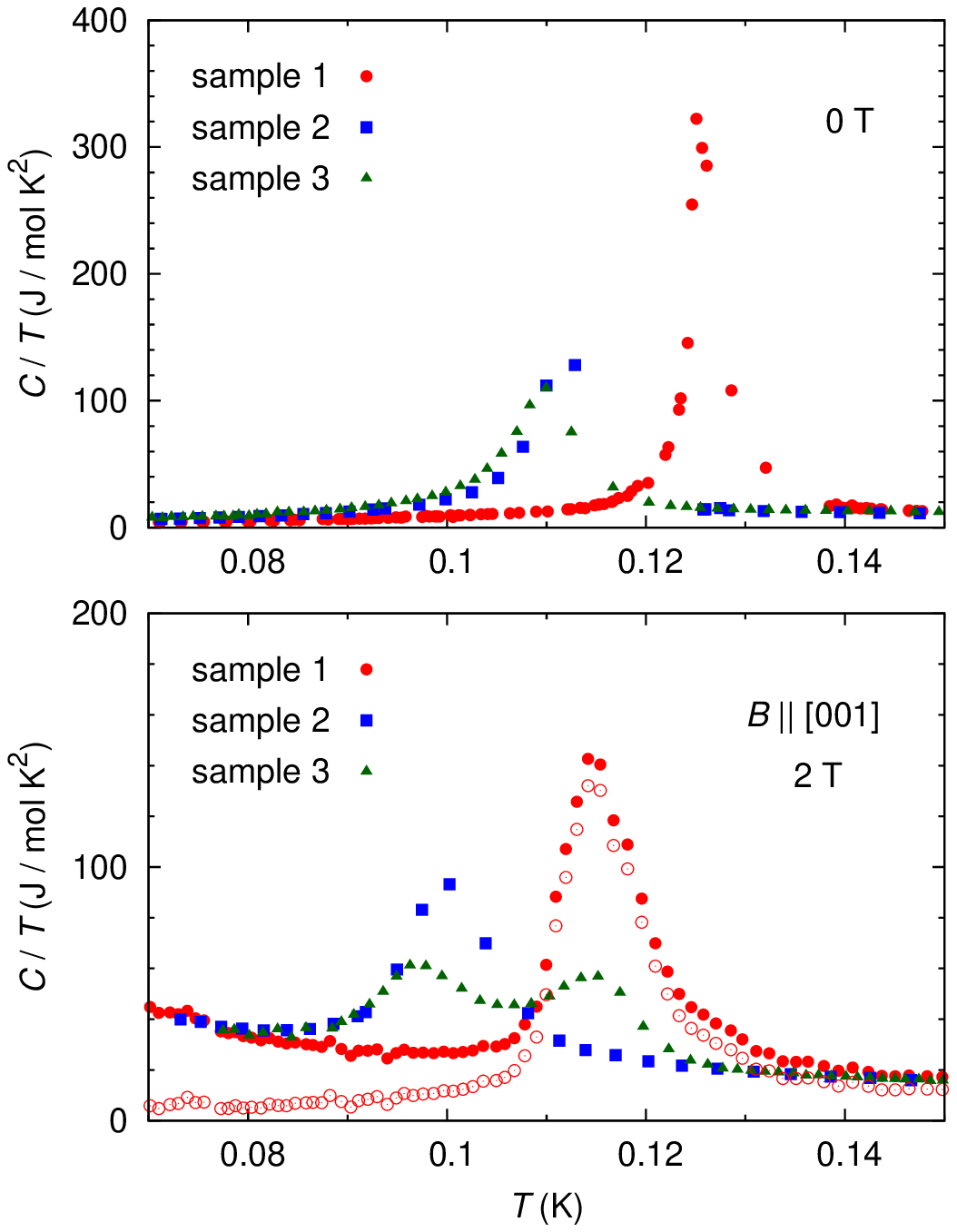} 
\end{center}
\caption{
Temperature dependences of the specific heat divided by temperature, $C/T$, for the samples 1, 2, and 3 at (a) 0 and (b) 2~T for $B \parallel [001]$.
The data of the samples 2 and 3 were taken from Refs.~\onlinecite{Onimaru2016PRB2} and \onlinecite{Onimaru2011PRL2}, respectively.
Open symbols in (b) represent the data after subtracting the nuclear contribution for the sample 1.
}
\label{sample}
\end{figure}

Figure~\ref{sample}(a) compares the specific-heat data of three samples of $\piz$ in zero field.
The data of the sample 1 were obtained in the present study whereas the data of the samples 2 and 3 were taken from Refs.~\onlinecite{Onimaru2016PRB2} and \onlinecite{Onimaru2011PRL2}, respectively.
It has been demonstrated that the antiferro-quadrupole (AFQ) transition temperature $T_{\rm Q}$ of $\piz$ depends on the sample quality.
Clearly, the present sample has the highest $T_{\rm Q}$ of 0.125~K.
Moreover, the peak height of the sample 1 is two times larger than the other two.
The sample 2 has $T_{\rm Q}$, almost the same as $T_{\rm Q}$ of the sample 3, but exhibits a specific-heat peak slightly larger than that of the sample 3.
Therefore, on the basis of $T_{\rm Q}$ and the specific-heat peak, 
it is possible to consider that the sample 1 has the highest quality among the three, and the sample 2 has slightly better quality than the sample 3,
although the residual resistivity ratio (RRR) of the sample 1 is comparable to that of the sample 2 (RRR $\sim 100$).

Figure \ref{sample}(b) compares the specific-heat data of the three samples at 2~T for $B \parallel [001]$.
Here, the nuclear specific heat $C_{\rm N}$ was not subtracted in the data represented by closed symbols.
The low-temperature upturn in $C(T)/T$ at 2~T disappears by subtracting the nuclear contribution, as exemplified in Fig.~\ref{sample}(b) by open circles for the sample 1.
Thus, the nuclear specific heat is not significantly large compared with the specific-heat anomaly associated with the AFQ transition. 
At 2~T, only the sample 3 shows a double transition, which remained observed in the relatively wide field range $1.5\ {\rm T} \lesssim B \lesssim 3$~T.\cite{Onimaru2011PRL2}
This study revealed that a double transition can also be seen for the sample 1, but in the narrower field range $2.2\ {\rm T} \lesssim B \lesssim 2.7$~T.
Therefore, at least for the samples 1 and 3, the small phase pocket exists, although the field range is slightly different.
The presence of the small phase pocket in the sample 2 is unclear because specific-heat measurements were not performed in the key field range $2\ {\rm T} < B < 3$~T.
In any case, the magnetic-field-induced small phase pocket depends on the sample quality.

\section*{II. Entropy analysis}

\begin{figure}
\begin{center}
\includegraphics[width=6.in]{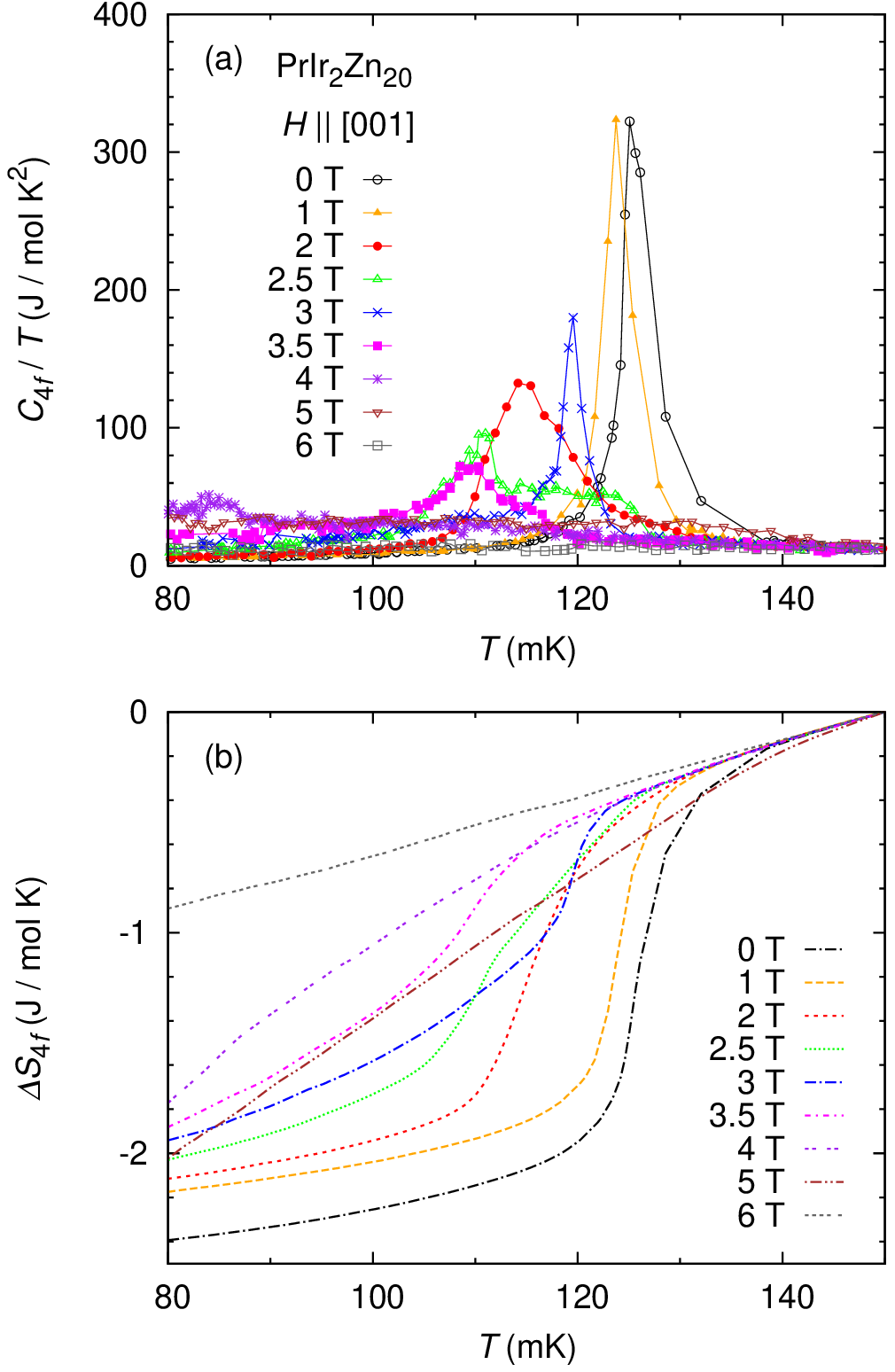} 
\end{center}
\caption{
Temperature dependences of (a) $C_{4f}/T$ and (b) $\Delta S_{4f}$ for the sample 1 at several magnetic fields for $B \parallel [001]$.
}
\label{B001}
\end{figure}

\begin{figure}
\begin{center}
\includegraphics[width=6.in]{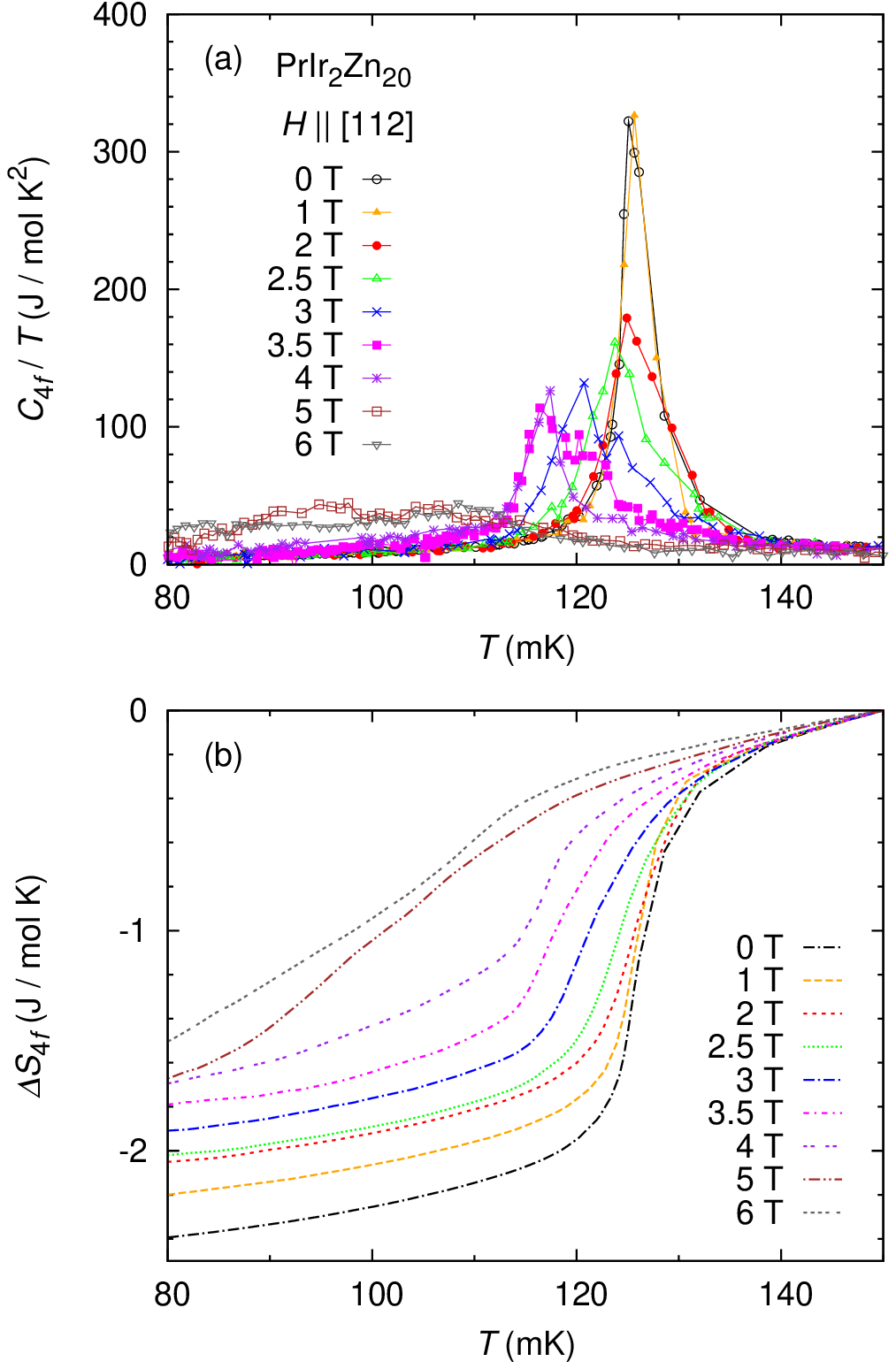} 
\end{center}
\caption{
Temperature dependences of (a) $C_{4f}/T$ and (b) $\Delta S_{4f}$ for the sample 1 at several magnetic fields for $B \parallel [112]$.
}
\label{B112}
\end{figure}

\begin{figure}
\begin{center}
\includegraphics[width=6.in]{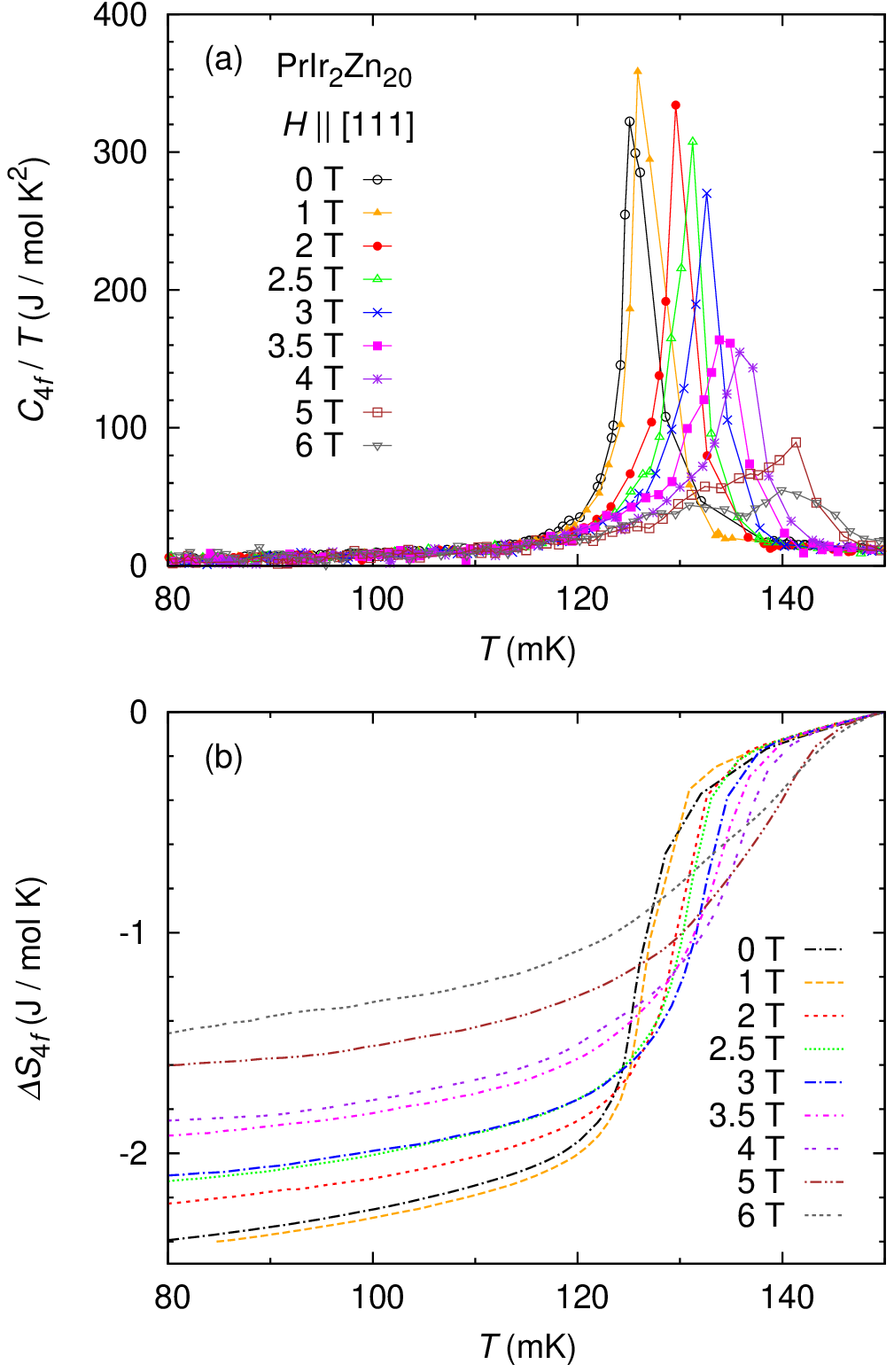} 
\end{center}
\caption{
Temperature dependences of (a) $C_{4f}/T$ and (b) $\Delta S_{4f}$ for the sample 1 at several magnetic fields for $B \parallel [111]$.
}
\label{B111}
\end{figure}

\begin{figure}
\begin{center}
\includegraphics[width=6.in]{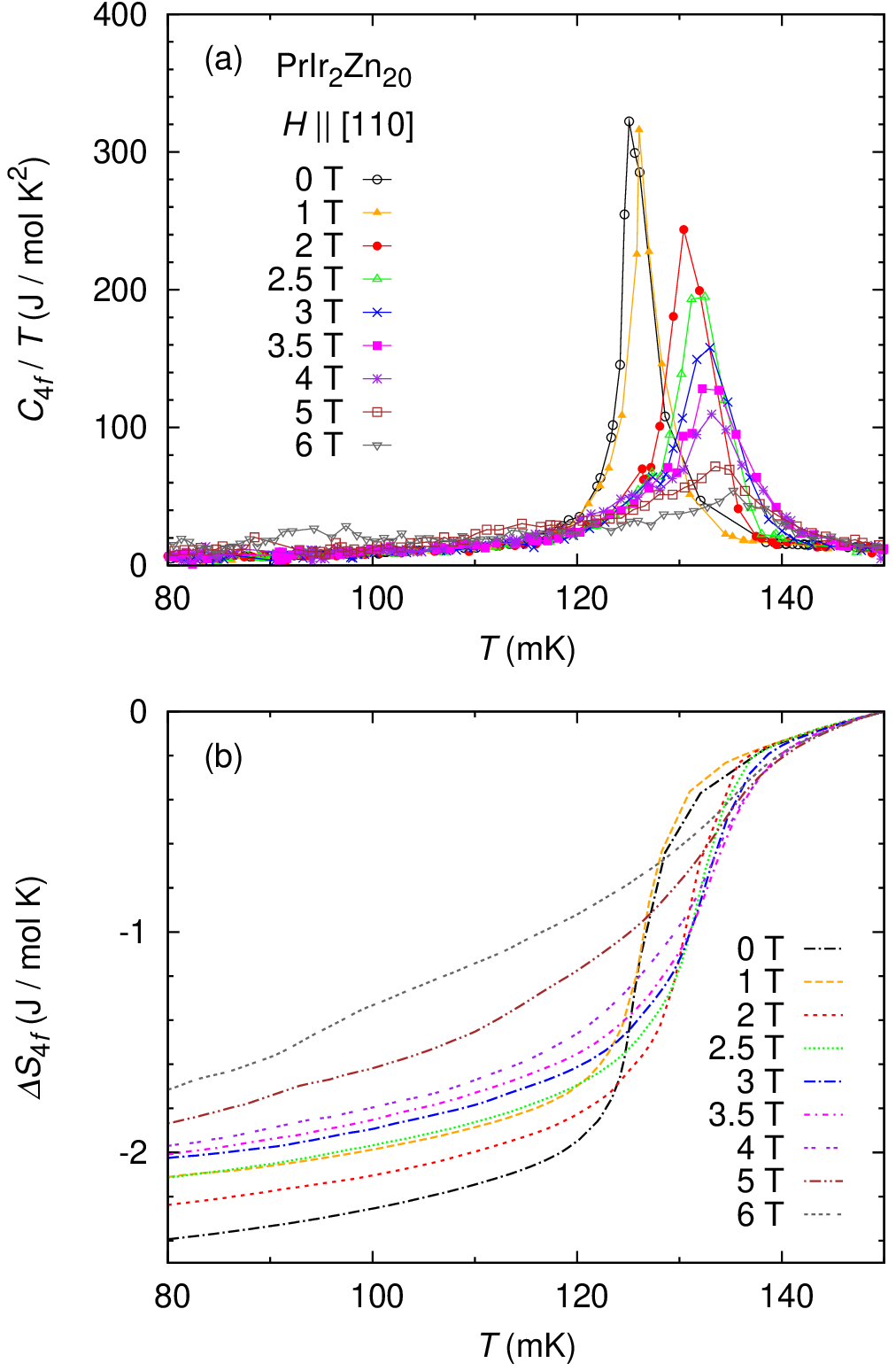} 
\end{center}
\caption{
Temperature dependences of (a) $C_{4f}/T$ and (b) $\Delta S_{4f}$ for the sample 1 at several magnetic fields for $B \parallel [110]$.
}
\label{B110}
\end{figure}

Figures \ref{B001}(a) shows the temperature dependence of the $4f$ contribution to the specific heat $C_{4f}$ ($=C-C_{\rm N}$) of the sample 1 divided by temperature, $C_{4f}/T$, at several magnetic fields for $B \parallel [001]$.
The relative change of the $4f$ contribution to the entropy can be estimated by integrating $C_{4f}/T$ as
\begin{equation}
\Delta S_{4f}(T)=S_{4f}(T)-S_{4f}(0.15\ {\rm K})=-\int_T^{0.15\ {\rm K}} \frac{C_{4f}(T)}{T}dT.
\end{equation}
Here, $\Delta S_{4f}$ is measured from the value at $T=0.15$~K. 
Temperature dependence of $\Delta S_{4f}$ for $B \parallel [001]$ is presented in Fig.~\ref{B001}(b).
In the same manner, temperature dependences of $C_{4f}/T$ and $\Delta S_{4f}$ for $B \parallel [112]$, $[111]$, and $[110]$ are shown in Figs.~\ref{B112}, \ref{B111}, and \ref{B110}, respectively.
In these figures, the entropy change occurs mainly due to the AFQ transition, except for the data above 4~T in $B \parallel [001]$,
at which the AFQ transition is significantly suppressed and a characteristic temperature $T_0$ is enhanced remarkably.

\clearpage

\clearpage

\clearpage

\clearpage

\clearpage

\end{document}